\documentclass[12pt,onecolumn,aps,pra,floatfix,superscriptaddress,preprint, showpacs,groupedaddress]{revtex4-1}

\usepackage{upgreek}
\usepackage{graphicx}
\usepackage{graphics}
\usepackage{amssymb}
\usepackage{amsmath}
\usepackage{epsfig}
\usepackage{latexsym}
\usepackage{color}
\usepackage{rotating}
\usepackage{subfigure}
\usepackage{float}
\usepackage[breaklinks=true]{hyperref}
\usepackage[labelfont=bf]{caption}
\usepackage{ragged2e}
\usepackage{braket}
\usepackage{physics}
\usepackage{mathtools}

\usepackage{hhline}
\usepackage{tabularx}
\usepackage{dcolumn}
\DeclareCaptionJustification{justified}{\justifying}
\hypersetup{
  colorlinks   = true, 
  urlcolor     = blue, 
  linkcolor    = blue, 
  citecolor   =  red 
}

\usepackage{soul} 

\begin{document}

\title{Nanomechanical test of quantum linearity}

\author{Stefan Forstner$^1$, Magdalena Zych$^1$, Sahar Basiri-Esfahani$^2$, Kiran E. Khosla$^3$, and Warwick P. Bowen$^1$\footnote{w.bowen@uq.edu.au}}
\affiliation{$^1$Australian Centre for Engineered Quantum Systems, School of Mathematics and Physics, The University of Queensland, Australia}
\affiliation{$^2$Department of Physics, Swansea University, Singleton Park, Swansea SA2 8PP, Wales, United Kingdom }
\affiliation{$^3$QOLS, Blackett Laboratory, Imperial College London, United Kingdom}

\begin{abstract}

Spontaneous wavefunction collapse theories provide the possibility to resolve the measurement problem of quantum mechanics. However, the best experimental tests have been limited by thermal fluctuations and have operated at frequencies far below those conjectured to allow the physical origins of collapse to be identified. Here we propose to use high-frequency nanomechanical resonators to surpass these limitations. 
We consider a specific implementation that uses a quantum optomechanical system cooled to near its motional ground state. The scheme combines phonon counting
with efficient mitigation of technical noise, including non-linear photon conversion and photon coincidence counting. 
It is capable of resolving the exquisitely small phonon fluxes required for a conclusive test of collapse models as well as potentially identifying their physical origin.

\end{abstract}

\maketitle

\section{Introduction}
Quantum mechanics is one of the most transformative physical theories of the 20th century. However, while the evolution of the quantum wave function is deterministically described by Sch\"odinger's equation, the outcome of a measurement is probabilistic, given by Born's rule. Despite recent progress \cite{Ozawa_Bayes_1997,Zurek_Darwinism_2009,Shrapnel_Born_2018}, there is no consensus on how to reconcile these two viewpoints, as illustrated by the measurement paradox \cite{Schroedinger_Cat_1935}. There are two conceptually distinct approaches: either the interpretative postulates must be modified \cite{Adler_Quantum_2009,Zeh_Measurement_1970,Everett_Relative_1957,Heisenberg_Prinzipien_1930,Bohm_Hidden_1952}, or quantum mechanics approximates a deeper theory yet to be discovered. The later approach gives rise to collapse models \cite{Bassi_Review_2012,Bassi_Gravitational_2017}, postulating a stochastic nonlinear modification to Schr\"odinger's equation. Irrespective of whether they successfully allow the reconciliation of quantum evolution and measurement theory, these collapse models are considered the only mathematically consistent, phenomenological modifications against which quantum theory can be tested \cite{Adler_Quantum_2009,Ferialdi_Dissipative_2012}. 

The most universal and well studied collapse model is Continuous Spontaneous Localization (CSL) \cite{Pearle_Localization_1989,Ghirardi_Markov_1990},
which serves as a framework to describe a variety of collapse mechanisms \cite{Bassi_Review_2012,Bassi_Gravitational_2017,Ghirardi_gravity_1990,Adler_Nonwhite_2007,Adler_Nonwhite_2008,Smirne_Dissipative_2014,Smirne_Dissipative_2015,Bassi_Breaking_2010}.
In CSL, a collapse noise field is introduced which couples nonlinearly to the local mass density. In its simplest form this noise is white, and the model has two parameters --- the collapse rate $\lambda_c$, which determines the interaction strength with the collapse noise field, and the correlation length $r_c$, which determines the spatial resolution of the collapse process \cite{Ghirardi_Markov_1990,Adler_Quantum_2009}.
The correlation length is expected to be $\sim100$~nm \cite{Adler_Bounds_2006}, since the behaviour of larger systems is generally adequately described by classical theories, whereas quantum mechanics appears to apply on smaller scales.
Refined dissipative and coloured models introduce two additional parameters, associating a temperature and high-frequency cut-off to the collapse noise field to ensure energy conservation and permit an identifiable physical origin of collapse \cite{Adler_Nonwhite_2007,Adler_Nonwhite_2008,Smirne_Dissipative_2014,Smirne_Dissipative_2015,Bassi_Breaking_2010}. 
Based on the assumption that the origin is of cosmological nature, and thermalised to the photon-, neutrino-, or gravitational wave background,
the high-frequency cut-off is estimated to occur at $\Omega_{\rm csl}/2\pi\sim 10^{10}-10^{11}$~Hz \cite{Bassi_Breaking_2010}.

To date, the most stringent unambiguous upper bounds on the collapse rate at the expected correlation length are based on mechanical resonators, with signatures of spontaneous collapse expected to manifest as an anomalous temperature increase.
However, the suggested lower bounds to the collapse-induced heating are lower than one phonon per day \cite{Adler_Bounds_2006,Bassi_Breaking_2010,Ghirardi_Unified_1986}.
The challenge of resolving these exquisitely small collapse signatures over a large thermal noise background has precluded conclusive tests of CSL, and has also introduced significant challenges for data interpretation \cite{Vinante_Improved_2017}.
Even were these issues resolved, quantum backaction heating \cite{Khalili_backaction_2012,Nimmrichter_Optomechanical_2014} would remain orders of magnitude larger than the predicted collapse signatures.
Moreover, with micron- \cite{Vinante_Cantilever_2015,Vinante_Improved_2017} to meter-sizes \cite{Carlesso_Gravitational_2016,Helou_LISA_2017}, the resonators employed to-date are larger than the anticipated correlation length and have frequencies far below the expected high-frequency cut-off. As such they are unable to provide insight into the physical origin of collapse \cite{Adler_Nonwhite_2007,Adler_Nonwhite_2008,Smirne_Dissipative_2014,Smirne_Dissipative_2015,Bassi_Breaking_2010}.

  \begin{figure}[h!]
  \includegraphics{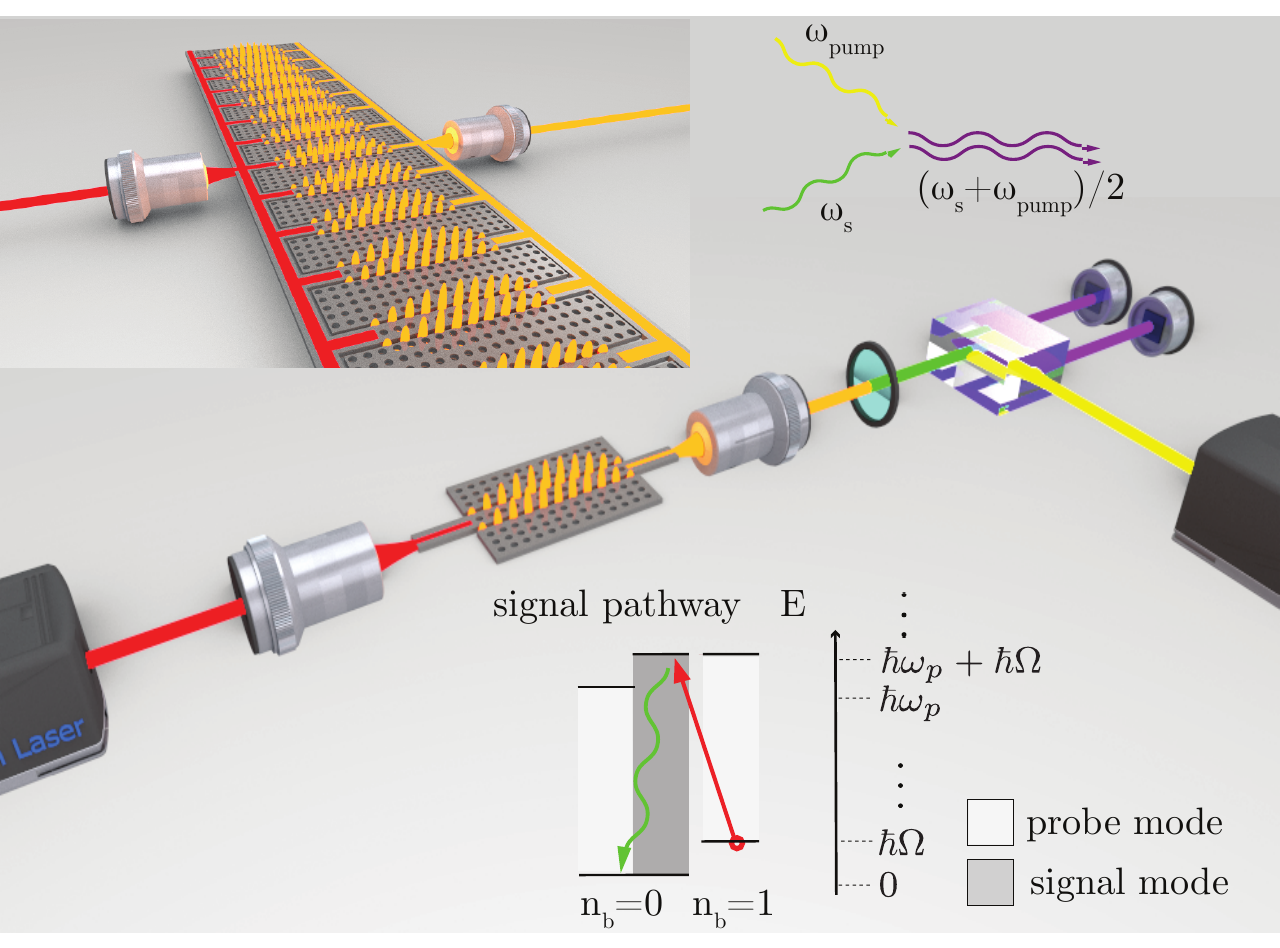}
\caption{{\bf Illustration of protocol.}
Top left: array of optomechanical cavities.
Top right: nonlinear pair production from a signal and pump photon (frequency $\omega_{\rm pump}$).
Bottom: Energy level diagram for scattering of a probe photon with a phonon.
$n_b$:  phonon number.
}
  \label{Fig_Setup}
    \end{figure}

In this work we propose to test collapse theories with high frequency nanomechanical resonators. This offers the advantages of miniaturisation to match the expected collapse correlation length, resonance frequency around the high-frequency cut-off, and the abilities to both exponentially suppress thermal phonons via passive cryogenic cooling and apply quantum measurement techniques to improve performance.
To assess the approach, we develop a specific experimental implementation that makes use of phonon counting in a nanoscale mechanical resonator.  
Our proposal includes new mitigation strategies for optical, thermal, electrical and quantum back-action noise that, for the first time, provide a way to bring each of these noise sources below the expected lower bounds for collapse-induced heating. We conclude that with challenging but plausible improvements in the state-of-the-art our approach could conclusively test CSL, closing the gap between measured upper bounds and predicted lower bounds on the collapse rate, and could also potentially identify the physical mechanism underlying the collapse process.
This provides an experimental pathway to answer one of the longest standing questions in physics, and  
also opens up possibilities for laboratory tests of astrophysical models of dark matter \cite{Riedel_undetectable_2013,Riedel_collisions_2017}, 
and other exotic particles \cite{Riedel_diffusion_2015}.

\section*{Results}

\subsection*{Basic protocol}

Our protocol is illustrated in Fig. \ref{Fig_Setup}, and is based on a gigahertz nanomechanical resonator, or array thereof,
within a millikelvin environment. 
As opposed to standard optomechanical measurement, consisting of an optical cavity linearly coupled to a mechanical resonator \cite{Aspelmeyer_Review_2014}, we propose to perform phonon counting in a three-mode optomechanical system where two optical modes
are coupled via a mechanical resonator with resonance frequency $\Omega$.
This allows collapse signatures to be spectrally distinguished from most noise sources.
One mode, the {\it probe mode}, is excited by a continuous weak laser at its resonance frequency $\omega_p$.
In the ideal case, the other, the {\it signal mode} at frequency $\omega_s=\omega_p+\Omega$, is only excited by resonant anti-Stokes Raman scattering between collapse induced phonons and probe photons.
A single-photon readout scheme minimises both absorption heating  \cite{Meenehan_millikelvin_2014} and quantum back-action heating.
Signal photons are spectrally separated from probe photons by a filter cavity, while dark counts are suppressed by nonlinearly downconverting signal photons to pairs and performing coincidence detection.

As a concrete example, we consider using a three-mode photonic-phononic crystal optomechanical system, such as proposed in \cite{Chang_Array_2011,BasiriEsfahani_Phonon_2012,Ludwig_TwoMode_2012,Safavi_Naeini_traveling_2011}. 
We choose most parameters based on those achieved in \cite{MacCabe_Ultralong_2019}, with a mechanical resonance frequency $\Omega/2\pi=5.3$~GHz, a mechanical damping rate $\Gamma/2\pi=108$~mHz, an effective mass $m_{\rm eff}=136$~fg, and thermalisation to the base temperature of a dilution refrigerator ($T=10$~mK).
We use the theoretical scattering-limited intrinsic decay rate of $\kappa_{p,0}=\kappa_{s,0}=2\pi\cdot 9.2$~MHz calculated for these devices \cite{Ren_Crystal_2019} for both optical modes, where the subscripts `$p$' and `$s$' distinguish the probe and signal mode throughout \cite{Aspelmeyer_Review_2014}.
Finally, we assume a tenfold improved single-photon optomechanical coupling rate of $g_0/2\pi=11.5$~MHz, as predicted to be feasible with optimized designs \cite{Matheny_coupling_2018}.

\subsection*{Phonon flux induced by CSL}

The CSL phonon flux is $\dot n_c= \lambda_c D$, where $D$ is a geometrical factor that quantifies the susceptibility of the resonator to spontaneous collapse.
The requirement that CSL should resolve the measurement problem introduces lower bounds on $\lambda_c$, and therefore on the phonon flux. 
Adler proposed  $\lambda_c\geq 10^{-8\pm 2}$~s$^{-1}$ from the postulate that collapse should account for latent image formation in photography \cite{Adler_Bounds_2006}, while Bassi {\it et al.} proposed $\lambda_c\geq 10^{-10\pm 2}$~s$^{-1}$ from the presumption that collapse should occur in the human eye \cite{Bassi_Breaking_2010}.
We estimate  $D=5.1\cdot 10^{5}$ for our proposed device \cite{Vinante_Cantilever_2015} (see Supplemental Material \cite{Supp}), which combined with these bounds implies minimum CSL induced phonon fluxes of $\dot n_c=5.1\cdot10^{-3\pm 2}$~s$^{-1}$ and $\dot n_c=5.1\cdot10^{-5\pm 2}$~s$^{-1}$, respectively.

\subsection*{Optomechanical dynamics and conversion efficiency}

If the oscillator is initially in its ground state,
with one photon in the probe mode, a phonon introduced by spontaneous collapse
prepares the state $\ket{n_b n_p  n_s}=\ket{110}$, where $n_b$ is the phonon number in the mechanical resonator, while $n_p$ and $n_s$ are the photon numbers in probe- and signal-mode, respectively. 
The optomechanical conversion efficiency $\eta_{\rm om}$ for this state to emit a signal photon at frequency $\omega_s$
is obtained by numerically solving the Born-Markov master equation taking into account that $\Gamma\ll\kappa_p,\kappa_s$ (see Methods).
We choose the external probe decay rate $\kappa_{p,\rm ex}/2\pi=2.2$~MHz \cite{Aspelmeyer_Review_2014}, allowing operation at the threshold of strong coupling with $g_0\approx\kappa_p$. 
This is advantageous for efficient conversion of collapse-induced phonons to signal photons and ensures low occupancy, minimising noise, as discussed later.
We choose the signal mode to be significantly overcoupled ($\kappa_{s,\rm ex}/2\pi=0.7\kappa_{s}=21$~MHz) in a trade-off between optimising the conversion efficiency and suppressing noise from direct occupancy of the signal mode (see later).
Together, these external decay rates result in relatively high conversion efficiency of $\eta_{\rm om}=0.32$.

 \subsection*{Noise sources}

\begin{figure}[!htbp]
	 \includegraphics{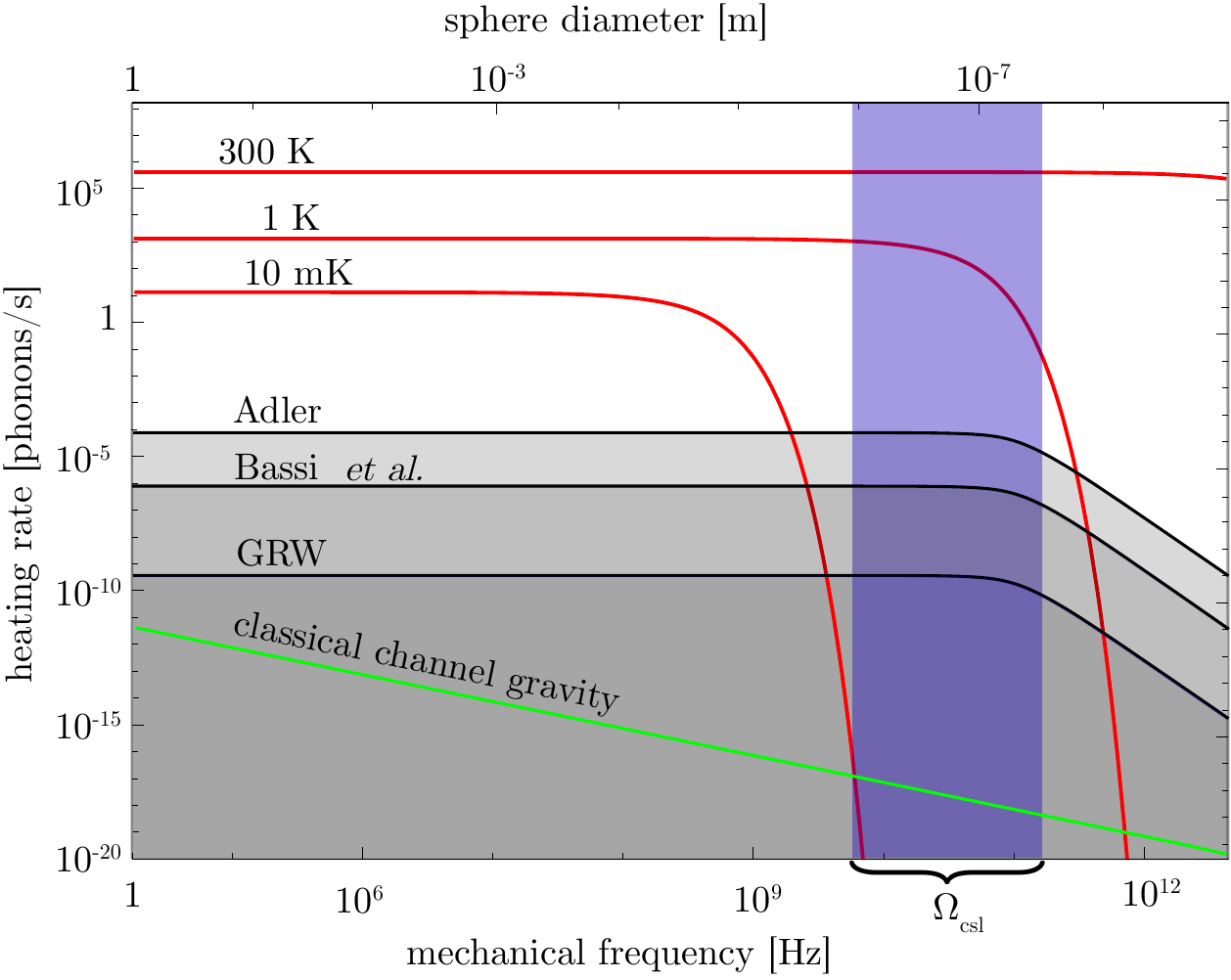}
\caption{{\bf Heating rates of a $Q=\Omega/\Gamma=10^7$ silica sphere resonator vs. mechanical frequency and sphere diameter}. Red traces: heating due to coupling to the thermal environment at temperatures 300,1 and 0.01 K. Gray shaded: Lower bounds on CSL heating rates for a sphere, according to  Adler \cite{Adler_Bounds_2006}, Bassi {\it et al.} \cite{Bassi_Breaking_2010} and GRW \cite{Ghirardi_Unified_1986}, assuming the fundamental mechanical breathing mode frequency $\Omega=c/R$, with sphere radius $R$ and speed of sound $c=3000$ m/s. CSL heating rates drop once the resonator becomes smaller than the noise correlation length, which is set to $r_c=10^{-7}$~m. Green: lower bound on heating rate predicted from classical channel gravity \cite{Kafri_Classical_2014,Khosla_Classical_2018,Altamirano__Pairwise_2018}. At high frequencies and low temperatures, collapse signatures exceed the thermal heating. Blue shaded: proposed range of $\Omega_{\rm csl}$.}
  \label{Fig_heating} 
\end{figure}

Four classes of noise can potentially imitate a collapse signal:  thermal phonons, probe photons that leak through the system, phonons introduced by the measurement process, and detector dark counts.
Photons that leak through the system can be efficiently filtered using a standard laser stabilisation reference cavity \cite{Kessler_Laser_2011}
(see Table \ref{tab:noise comparison} and Methods), and will not be considered further here.

{\it Thermal phonons.}
 A collapse signature is resolvable in a thermal noise background if $\dot n_c/\dot n_{\rm th}>1$, where $\dot n_{\rm th}=\Gamma (e^{\hbar \Omega/k_BT}-1)^{-1}$ is the thermal phonon flux.
This gives a minimum testable collapse rate $\lambda_{c,\rm th}=\dot n_{\rm th}/D$. 
Existing experiments have operated with comparatively low frequency oscillators in the high temperature limit, $k_BT \gg \hbar\Omega$ \cite{Vinante_Cantilever_2015,Vinante_Improved_2017,Carlesso_Gravitational_2016,Helou_LISA_2017}, with thermal phonon flux significantly larger than Bassi {\it et al.}'s lower bound,
and have sought to resolve small collapse signatures on top of this large thermal noise background.
A significant advantage of our approach is that miniaturisation and cryogenic cooling allow access to the regime where $k_BT \ll \hbar\Omega$.
The average thermal phonon occupation is then exponentially suppressed due to Bose-statistics, $\dot{n}_{\rm th} \approx \Gamma e^{-\frac{\hbar\Omega}{k_ BT}}$.
Fig. \ref{Fig_heating} shows this exponential suppression as a function of resonator size, and in comparison to the CSL signal, for the simple example of the fundamental breathing mode of a silica sphere (see Supplemental Material \cite{Supp} for calculation).
As can be seen, for gigahertz resonators at millikelvin temperatures the exponential suppression allows thermal phonon fluxes beneath both Adler and Bassi {\it et al.}'s lower bounds.
For the proposed photonic crystal device, we find $\lambda_{c,\rm th}=1.2\cdot 10^{-17}$~s$^{-1}$,
also well beneath both bounds.
This provides the potential for
unambiguous tests of collapse models.

  \begin{figure}[!htbp]
\includegraphics{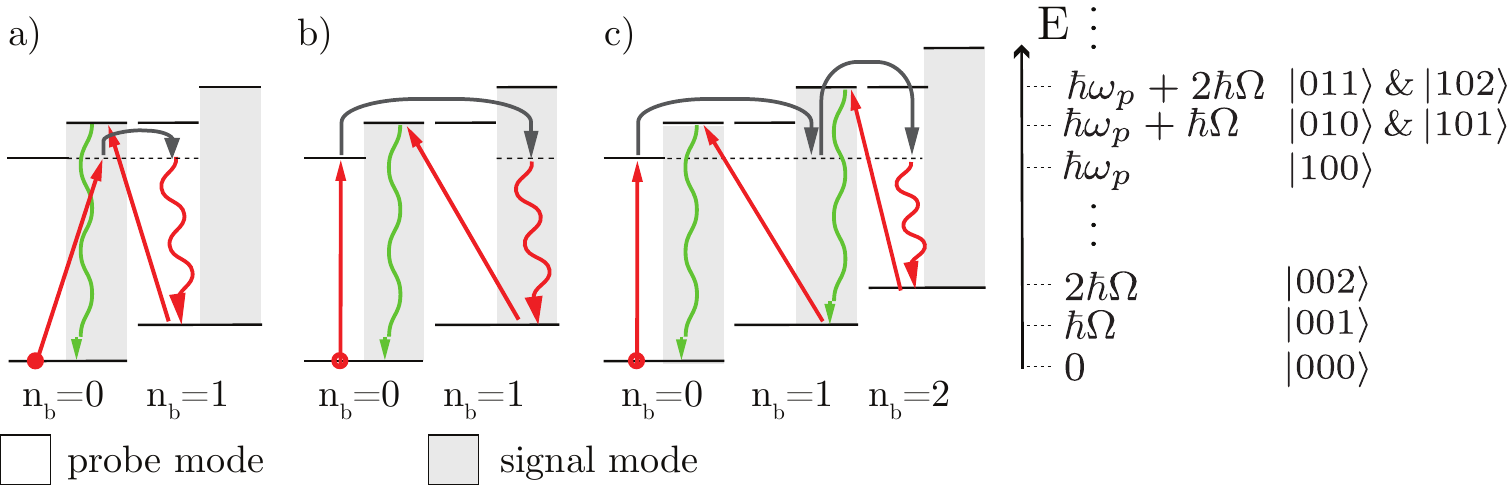}
\caption{{\bf Signal pathways due to measurement-induced phonons.}
a) Phonon created after direct excitation of signal mode.
b) Phonon due to counter-rotating transition.
c) Two phonons created by counter-rotating transition followed by resonant transition. 
}
\label{Fig_Energy}
      \end{figure}

\begin{figure}[!htbp]
	\includegraphics{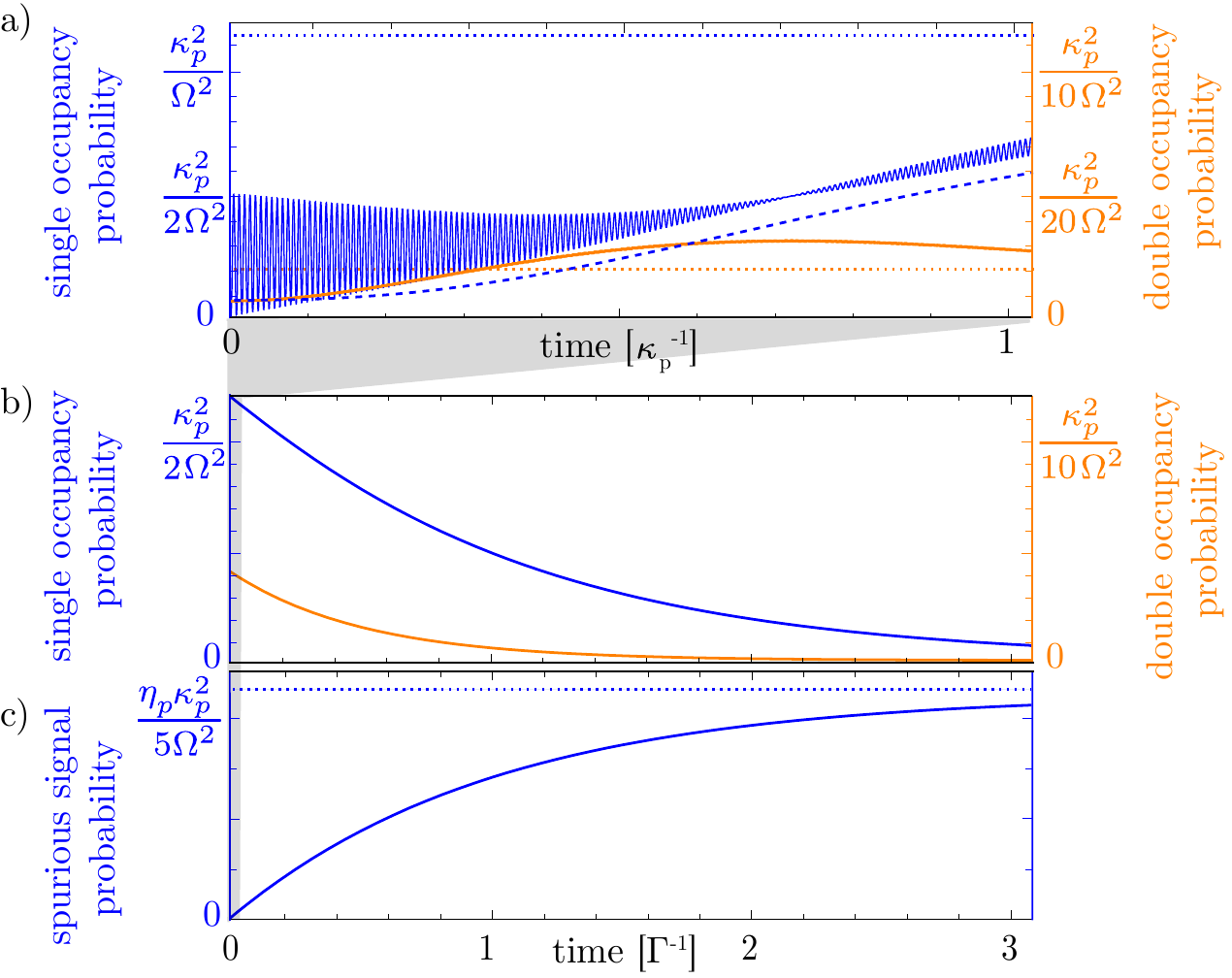}%

\caption{{\bf Numerical calculations of noise magnitude.}
a) Blue (orange): occupancy of density matrix elements containing one (two) phonon(s);
dashed blue: contribution from direct signal mode excitation.
Fast oscillations on timescale $\Omega^{-1}$ correspond to the counter-rotating transition $\ket{010}\leftrightarrow\ket{101}$; 
slow oscillations to the resonant process $\ket{101}\leftrightarrow\ket{210}$, with period $g_0^{-1}=\kappa_p^{-1}$. 
Dotted lines: asymptotic values for $\kappa_p^{-1},\kappa_s^{-1}\ll t\ll\Gamma^{-1}$.
b) Same as (a) for $t\sim\Gamma^{-1}\gg\kappa_p^{-1}, \kappa_s^{-1}$. 
c) Cumulative probability $p_{\rm om}(t)$ of a probe photon creating a photon at frequency $\omega_s$. 
Dotted line: asymptotic value for $t\gg\Gamma^{-1}$.
}
\label{figSOM}
\end{figure}

{\it Measurement-induced phonons.}
Phonons introduced by the optomechanical measurement can imitate collapse signatures. These phonons are created by non-resonant scattering processes between the signal and probe modes, the three lowest order of which are represented in Fig. \ref{Fig_Energy}.
We calculate the probability of phonon occupancy due to these processes numerically by solving the Born-Markov master equation (see Methods).
We find that each of these processes is suppressed by the square of the resolved sideband ratio $\Omega/\kappa_p$, with predicted phonon occupancies shown in Fig. \ref{figSOM} (a) and (b).
Photoabsorptive heating can also introduce phonons. 
However, it only adds a negligible contribution to the measurement-induced phonon occupancy (see Table \ref{tab:noise comparison} and Methods).


A measurement-induced phonon can only be converted to a collapse-imitating photon at frequency $\omega_s$ if it scatters with a second photon entering the probe mode within the lifetime $\Gamma^{-1}$ of the mechanical resonator (see Fig. \ref{Fig_Energy}).
It is therefore possible to suppress these photons by operating with a low average photon occupancy $\bar n_p$. Here, we choose the photon occupancy so that the probability of a photon entering the probe mode during one mechanical oscillation lifetime is $\eta_p = \bar n_p \kappa_p/\Gamma\sim$~1\%.
This reduces the rate of measurement-induced photons by a factor of a hundred. 
The cumulative probability of a probe photon generating a phonon, and a second probe photon then causing emission of a photon at frequency $\omega_s$, is shown in Fig. \ref{figSOM} (c).
The asymptotic probability is $p_{\rm om}(t\rightarrow \infty)=8.4\cdot 10^{-8}$ (see Supplemental Material \cite{Supp}).

{\it Coincidence dark counts.}
 Detecting collapse induced phonons at the predicted rate of less than one per day necessitates very high suppression of photon dark counts, which typically occur at hertz to kilohertz rates.
One way to  achieve this is to nonlinearly downconvert signal photons to pairs (bottom right inset, Fig. \ref{Fig_Setup}) 
using a bright pump beam in a third-order nonlinear medium.
It has been shown that this process can convert single photons to pairs with near-unit efficiency $\eta_\chi$ \cite{Langford_Conversion_2011} (see Supplemental Material \cite{Supp}). 
A signal is recorded only if a coincidence detection event is registered.
The coincidence dark count rate is suppressed as the square of the single-detector dark count rate $R_{d,1}$, $R_{d,2}=R_{d,1}^2 \cdot \tau_c$, where $\tau_c$ is the coincidence timing resolution. 
For commercially available photon counters with $R_{d,1}=3.5$~s$^{-1}$ and $\tau_c=30$~ps \cite{Photonspot_private}, we predict $R_{d,2}\sim 3.7\cdot10^{-10}$~s$^{-1}$. 

 \subsection*{Minimum testable collapse rate}
 
 For $r_c=10^{-7}$~m, the rate of coincidence counts attributed to collapse is  $R_c=\lambda_c D \eta=5.5\cdot 10^{2}\lambda_c$, where the efficiency  $\eta=\eta_p\eta_{\rm om}\eta_{\chi}\eta_d\eta_{\rm f}=1.1\cdot10^{-3}$ quantifies the fraction of phonons in the mechanical resonator that result in a coincidence count, $\eta_{\chi}=0.95$, and $\eta_d=0.64$ is the coincidence detection efficiency (see Supplemental Material \cite{Supp}).
This rate must exceed the sum of the noise rates, setting the limit to the minimum observable collapse rate $\lambda_c$.
For optomechanically induced phonons, probe photons leaking through the system, and thermal phonons, $R_{\rm om}=\kappa_{p,\rm ex} \bar n_pp_{\rm om}(t\rightarrow \infty)\eta_f\eta_\chi\eta_d$, $R_{\rm phot}=\kappa_{p,\rm ex} \bar n_pp_f\eta_\chi\eta_d$, and $R_{\rm th}=\eta\dot n_{\rm th}$, respectively, where $\eta_f=0.56$ is the transduction efficiency through the filter and $p_f=3.5\cdot 10^{-10}$ the probability of a probe photon leaking through the filter (see Supplemental Material \cite{Supp}).
The numerical values and corresponding minimum testable collapse rates are given in Table \ref{tab:noise comparison} (see also Supplemental Material \cite{Supp}).
Optomechanical measurement-induced phonons and coincidence dark counts set comparable limits on $\lambda_c$, with negligible contributions from leaked probe photons, photoabsorption, and thermally excited phonons.  
The minimum testable collapse rate limited by all noise sources is $\lambda_c=\sum_i\lambda_{\rm c,i}=1.0\cdot10^{-12}$~s$^{-1}$, sufficient to test both Bassi {\it et al.}'s and Adler's proposals.

\begin{table}
\center
\begin{tabular}{c | c | c | c} 

noise type & scaling & signal rate~[s$^{-1}$]   &$\lambda_{c,\rm min}$~[s$^{-1}$] \\

\hhline{=|=|=|=}
\rule{0pt}{9pt} collapse noise & $\eta l^2\rho^2x_0^2$\cite{Supp} & $5.5\cdot 10^{2}\lambda_c$ & --- \\
\hline
\rule{0pt}{12pt} thermal & $e^{-\frac{\hbar\Omega}{k_BT}}$ &  $6.7 \cdot 10^{-15}$ & $1.2\cdot 10^{-17}$ \\
\hline
\rule{0pt}{10pt} optom. phonons &  $\eta_p\kappa^2/\Omega^2$ & $1.9 \cdot 10^{-10}$  & $3.5 \cdot 10^{-13}$  \\
\hline
\rule{0pt}{10pt} abs. heating & $l/(m^{\frac{1}{4}}\Omega)$ \cite{Supp} & $1.4 \cdot 10^{-14}$  & $ 2.6 \cdot 10^{-17}$ \\ %
\hline
\rule{0pt}{10pt} probe photons & see \cite{Supp}& $1.4\cdot 10^{-12}$  &$2.6\cdot 10^{-15}$\\
\hline
\rule{0pt}{9pt} dark counts & $R_{d,1}^2 \cdot \tau_c/N$ &$3.7\cdot 10^{-10}$ & $ 6.7\cdot 10^{-13}/N$ \\
\hline
\rule{0pt}{9pt} all noise & --- & $5.6\cdot 10^{-10} $ & $ 1.0\cdot 10^{-12}$ \\
\hline
\end{tabular}
\caption{{\bf Comparison of noise sources and respective testable CSL parameter $\lambda_c$.} $m$ is the oscillator mass, $l$ its linear size, $\rho$ its density, and $x_0$ its zero point motion.}
\label{tab:noise comparison}
\end{table}

 \subsection*{Signal rate and measurement time}
The predicted average time required to observe one signal due to CSL-collapse is $t_{\rm meas}=(\lambda_c D\eta)^{-1}$. 
Fully probing both Adler's and Bassi {\it et al.}'s proposals with a single optomechanical resonator, including their respective uncertainties, would require $t_{\rm meas}>57$~years.
Fabricating an array of $N$ optomechanical cavities on a silicon wafer \cite{Bekker_tuning_2018,Xu_cascaded_2006,GilSantos_cascaded_2016,Zhang_synchronisation_2011} (see Fig. \ref{Fig_Setup}), coupled to a single filter cavity, nonlinear medium and detector  (or a small number of such elements)
could significantly reduce this time to $t_{\rm meas}^{(N)}=t_{\rm meas}/N$, and also the dark count-limited testable collapse rate to $\lambda_{c,\rm det}^{(N)}=\lambda_{c,\rm det}/N$.
We estimate that $N\sim10^{4}$ may be feasible (see Supplemental Material \cite{Supp}), essentially eliminating detector dark counts as a limit, and allowing a reduction of the measurement time to about two days.

 \subsection*{Feasibility and alternative parameter regimes}
The only optomechanical parameters that must be improved from the current state-of-the-art \cite{MacCabe_Ultralong_2019} to realise our protocol, are a reduced optical linewidth (by a factor of $\sim50$), as predicted by theoretical modelling based on the device realized in \cite{Ren_Crystal_2019}, and an enhanced single-photon coupling rate (by a factor of $\sim10$), based on theoretical modelling in \cite{Matheny_coupling_2018}. 
Alternatively, effective enhanced single-photon coupling could be achieved by coupling to a qubit or other highly nonlinear system, e.g. as demonstrated in \cite{Pirkkalainen_hybrid_2013}.
Given the trajectory of the field, 
we estimate these requirements to be likely achievable in the intermediate future. 
Nevertheless, it is also useful to consider alternative realizations of the method.

{\it Quadratic coupling.}
While here we consider phonon-counting via an optomechanical Raman interaction,
in principle the method could be implemented with any low-noise phonon-counting method applied to a high-frequency oscillator \cite{Oconnell_quantum_2010,Dellantonio_nondemolition_2018,Sletten_Fock_2019,Cohen_counting_2014,ArrangoizArriola_nanomechanical_2019}. 
One promising approach may be quantum non-demolition measurement of phonon number using non-linear optomechanics \cite{Thompson_strong_2008}. 
In the regime of quadratic optomechanical coupling and resolved mechanical sidebands \cite{Aspelmeyer_Review_2014}, a collapse-induced phonon imparts a frequency shift $2\bar n_{\rm cav}^{1/2}g_0^{(2)}$ on the optical resonance at frequency $\omega$, where $\bar n_{\rm cav}=\langle a^\dagger a \rangle$ is the average intracavity photon number with $a$ the annihilation operator for the optical cavity field, and $g_0^{(2)}$ the zero-point quadratic coupling rate \cite{Thompson_strong_2008,Paraiso_Squared_2015,Hauer_nondemolition_2018}.
The shift is detectable if it is larger than the significant noise sources, which are random fluctuations in the probe frequency, absorption heating, and quantum-backaction from spurious linear coupling.

Considering Bassi {\it et al.}'s proposed mechanism \cite{Bassi_Breaking_2010}, taking $\bar n_{\rm cav}=10^2$ and assuming that the probe is shot noise limited, we find that a zero-point quadratic coupling rate of  $g_0^{(2)}\gtrsim3.5\sqrt{\kappa\Gamma}\bar n_{\rm cav}^{-3/2}\gtrsim 2\pi\cdot 28$~Hz would be sufficient for
the weakest possible collapse signal to exceed the probe frequency noise using the photonic-phononic crystal considered in the protocol above \cite{MacCabe_Ultralong_2019}
(see Methods). 
This is well within experimentally achieved values in optomechanical photonic crystals (e.g. $g_0^{(2)}/2\pi=245$~Hz in \cite{Paraiso_Squared_2015}).

Perhaps the most significant challenge in this approach would be to engineer a strong suppression of linear optomechanical coupling, so that the phonon flux due to quantum back-action does not exceed the predicted CSL signature. 
If using standard architectures, there is a fundamental limit to this suppression of linear coupling \cite{Miao_Limit_2009}. Hence, either a different architecture would have to be employed \cite{Kaviani_paddle_2015,Dellantonio_nondemolition_2018,Hauer_nondemolition_2018}, or the substantially more stringent condition $g_0^{(2)}\geq\kappa$ would have to be realized.
The phonon flux due to quantum back-action is given by  $\dot{n}_{\rm ba}=4g_0^2\bar n_{\rm cav}/\kappa=4g^2/\kappa$. 
To resolve a potential CSL signature, $\dot{n}_{\rm c}$ must be greater than $\dot{n}_{\rm ba}$.
As a result, the linear optomechanical coupling would need to be suppressed to $g\leq\sqrt{\lambda_c D \kappa /4}$. To test $\lambda_c=10^{-12}$, we find the condition $g/2\pi\lesssim 10^{-1}$~Hz, about seven orders of magnitude lower than typical linear coupling rates in photonic-phononic crystal structures  \cite{MacCabe_Ultralong_2019}.
While some architectures may in principle allow for vanishing linear coupling $g$, achieving the required suppression in practice may be challenging  \cite{Dellantonio_nondemolition_2018,Kaviani_paddle_2015,Anetsberger_Near-Field_2009}. 

In continuous operation, with currently available technology \cite{MacCabe_Ultralong_2019}, absorption heating would exceed the expected heating from collapse by about seven orders of magnitude.
Even with this very large heating in the continuous domain, it may be possible to resolve the problem by operating in a pulsed regime, so that each optomechanical measurement process is completed in a timescale much shorter than the time required for absorption events to create phonons.
In this case, the measurements would need to be sufficiently temporally spaced to allow for phonons to fully dissipate.

\subsection*{Discussion}

\begin{figure}[!htbp]
  \includegraphics{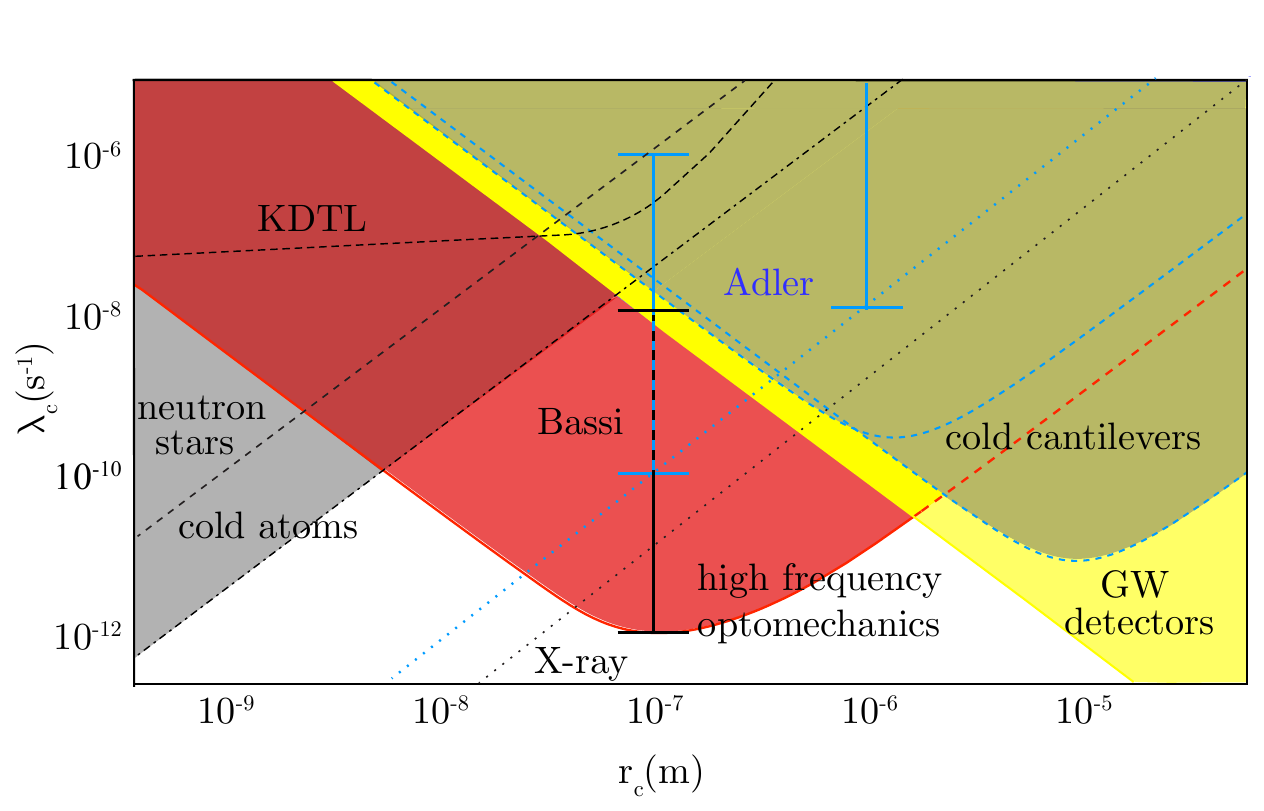}
\caption{{\bf Parameter diagram for CSL-model.}
Excluded upper bounds: gravitational wave detectors (yellow shaded); cold atoms (gray shaded); microcantilevers (dashed blue line); KDTL-interferometry (dashed black); Excluded for simple CSL only: neutron stars (dashed black) and X-ray (dotted black).
Proposed lower bounds: Adler (vertical blue bars and dotted blue line) and Bassi {\it et al.} (vertical black bar).
Red: predicted testable parameter space using our protocol. 
}

\label{figCSL}
\label{stats}
\end{figure}

Fig. \ref{figCSL} compares the predicted upper bound on the collapse rate $\lambda_c$ from our protocol to those of existing experiments,
together with Adler's and Bassi {\it et al.}'s lower bounds, and their uncertainties.
Existing upper bounds  are provided by the motional stability of gravitational wave interferometers  \cite{Carlesso_Gravitational_2016,Helou_LISA_2017,Nobakht_Unraveling_2018} (yellow region); the thermalisation of ultracold cantilevers \cite{Vinante_Cantilever_2015,Vinante_Improved_2017} (blue outlined region); Kapitza-Dirac-Talbot-Lau (KDTL)-Interferometry \cite{Fein_25kDa_2019,Nimmrichter_Matterwave_2011,Toros_Colored_2017,Feldmann_Parameter_2012} (dashed black); spontaneous X-ray emission from Germanium \cite{Curceanu_Xray_2016,Curceanu_LNGS_2020,Fu_Radiation_1997} (dotted black) and the observed temperature of neutron stars \cite{Adler_Fermi_2019,Stace_Neutron_2019} (dashed black), which are valid however only for white noise CSL; and cold atom interference \cite{Toros_Colored_2017} (gray region), though we note the controversy \cite{StamperKurn_Comment_2016} on the actual size of the superposition reported in \cite{Kovachy_Superposition_2015}.

The red shaded region in Fig. \ref{figCSL} could be tested by our protocol
as discussed above. In the case of white-noise collapse, the protocol could for the first time fully test Bassi {\it et al.}'s proposal. 
If collapse noise has one of the proposed physical origins \cite{Bassi_Breaking_2010,Smirne_Dissipative_2014,Smirne_Dissipative_2015}, the envisaged protocol would also for the first time probe Adler's prediction, which is in this case not tested by X-ray emission (black dotted line in Fig. \ref{figCSL}).
The resonance frequency is close to the frequency range in which a drastic frequency-dependent reduction of the collapse noise stemming from a physical origin is expected \cite{Bassi_Breaking_2010}. 
To identify the physical origin of collapse, and to differentiate between collapse-induced signal and technical noise, we suggest employing a number of mechanical resonators of slightly different frequencies, or one frequency-tunable resonator \cite{Pfeifer_tunable_2016}, at frequencies around $\Omega/2\pi\sim10$~GHz, such as reported in \cite{Ren_Crystal_2019}.

We also evaluate the capability of the protocol to constrain parameters in gravitational collapse models.
While for the Di\'osi-Penrose model \cite{Diosi_Gravitation_1984,Diosi_Models_1989,Penrose_Reduction_1996} we find that it cannot exceed existing bounds, for the classical channel gravity model in a typical parameter range \cite{Altamirano__Pairwise_2018,Kafri_Classical_2014,Khosla_Classical_2018} we predict about a one order-of-magnitude stronger bound than previously achieved \cite{tbp}  (see Supplemental Material \cite{Supp}).

In summary, we have proposed the concept of testing quantum linearity using high-frequency mechanical oscillators. 
This offers the advantages of thermal noise suppression to well below expected collapse signatures, and the potential for identification of the physical origin of collapse. 
As a possible implementation we suggest a protocol based on a dual-cavity high-frequency optomechanical device passively ground-state-cooled and operating in the strong coupling regime. This design, combined with nonlinear optical techniques to reduce dark counts, is predicted to allow measurement of the minuscule phonon-flux generated by collapse-induced heating.  While challenging, the protocol has the potential to conclusively test CSL, and thus whether collapse mechanisms can be invoked to resolve the measurement paradox. Unlike previous proposals and experiments, it is designed to allow for identification of the physical noise field underlying CSL, and for differentiation between excess technical noise and signatures of collapse.

\section*{Methods}  

\subsection*{Born-Markov master equation}  

To model the dynamics of the three-mode optomechanical system we employ the Born-Markov framework for open quantum systems \cite{Nunnenkamp_Single_2011,Liu_Strong_2013,BasiriEsfahani_control_2016}.
 The interaction picture Hamiltonian for our system is \cite{BasiriEsfahani_Phonon_2012,BasiriEsfahani_control_2016,Chang_Array_2011}
\begin{equation}
H_{\rm int}=\hbar g_0(b^{\dagger}e^{-i\Omega t}+be^{i\Omega t})(a_p^{\dagger}a_se^{i\Omega t}+a_pa_s^{\dagger}e^{-i\Omega t})+\hbar\sqrt{\kappa_{p,\rm ex}}(a_p^{\dagger}a_{\rm in}+a_{\rm in}^{\dagger}a_p),
\end{equation}
where $b$, $a_p$ and $a_s$ are annihilation operators for the mechanical mode and optical modes, respectively, and $a_{\rm in}$ is the coherent input field. The first term describes the mechanically mediated cross-coupling of the optical modes, while the second term describes the coherent excitation \cite{Aspelmeyer_Review_2014}. In the parameter regime of this work where $g_0\ll \Omega$ and $\Gamma \ll \kappa_p,\kappa_s,g_0$, the dynamics of the system can be described by the
Born-Markov master equation as \cite{Nunnenkamp_Single_2011,Liu_Strong_2013}
\begin{equation}
\frac{d\hat \rho}{dt}=-\frac{i}{\hbar}\big[H_{\rm int},\hat \rho\big] + \kappa_p \mathcal{D}[a_p]\hat \rho + \kappa_s \mathcal{D}[a_s]\hat \rho + \Gamma(1+\bar{n}_{\rm th}) \mathcal{D}[b]\hat \rho + (\Gamma \bar{n}_{\rm th} + \dot n_c)\mathcal{D}[b^{\dagger}]\hat \rho,
\label{eq: master_equation}
\end{equation}
where $\hat \rho$ is the density matrix, $\bar n_{\rm th}$ the mechanical mean thermal occupancy, 
and $\mathcal{D}$ the dissipating superoperator, $\mathcal{D}[A]\hat \rho=A\hat \rho A^{\dagger}-\frac{1}{2}(A^{\dagger}A\hat\rho+\hat\rho A^{\dagger}A)$. 
A weak phonon flux due to spontaneous collapse is described by $\dot n_c=\lambda_c D$, independent of its origin.
This allows us to model the conversion of a signal phonon to a signal photon, as well as creation of noise phonons introduced by measurement (see Supplemental Material \cite{Supp}).




\subsection*{Negligible sources of noise}

{\it Probe photons leaking though the system.}
 Probe photons passing directly from the laser through the optomechanical system, without a scattering event, could in principle imitate a signal, obfuscating collapse signatures.
We find that, using a standard laser stabilisation reference cavity as a filter \cite{Kessler_Laser_2011}, this noise can be suppressed well below both Adler's and Bassi {\it et al.}'s lower bounds.
Similarly, if a photon is created in an optomechanical conversion process and subsequently outcoupled into the signal mode, due to energy conservation it either remains at frequency $\omega_p$, or has a frequency reduced by integer multiples $n$ of the mechanical resonance frequency, $\omega_p-n\Omega$.  In both cases, this noise is doubly suppressed --- first by the suppression of the direct occupation pathway, and second by the filter. 
This makes probe photons that leak through the system a negligible source of noise (see Supplemental Material \cite{Supp} for details).

{\it Optical absorption heating.}
To estimate the phonon occupancy due to optical absorption,
we use the model for absorption heating in silicon optomechanical crystals outlined in \cite{Meenehan_millikelvin_2014,Ren_Crystal_2019}.
Photoabsorption creates an electronic excitation, which is then transferred to terahertz-frequency phonons. While radiating from the resonator to the environment with a geometry- and material-dependent rate $\gamma_{\rm THz}$, they also couple to lower energy phonons with a generally longer timescale, potentially exciting the mechanical resonator.  
In \cite{Meenehan_millikelvin_2014,Ren_Crystal_2019}, the average phonon number $\bar n_b$ is related to the average intracavity photon number $\bar n_{\rm cav}$ via $\bar n_b\propto \bar n_{\rm cav}^{1/3}$. 
We expect this relationship to break down when the time between photoabsorption events is long enough for the generated heat to fully dissipate, $\bar n_{\rm cav} \cdot \gamma_{\rm THz}/\kappa \lesssim1$, where $\kappa$ is the loaded optical decay rate, as any discrete photon absorption event is expected to create a fixed amount of heat.
In this case $\bar n_{\rm cav}$ determines the frequency of these events, but not the magnitude of dissipated heat.
We compute the average phonon number excited by of one probe photon in the mechanical resonator, $\bar n_{\gamma}$, due to photoabsorption for time $t_{\rm abs}$, at which the oscillator is in thermal equilibrium with the material, but not yet with the environment, $\Gamma^{-1}\gg t_{\rm abs} \gg \gamma_{\rm THz}^{-1}$.
For the proposed setup we find $p_{\rm abs}(t\rightarrow\infty)=6.1\cdot 10^{-12}$ and  $R_{\rm abs}=1.4\cdot10^{-14}$~s$^{-1}$ (see Supplemental Material \cite{Supp} for calculation details).

\subsection*{Measurement-induced phonons.}
A probe photon can create a noise phonon by coupling directly into the signal mode instead of the probe mode (Fig. \ref{Fig_Energy} (a)). 
This process is suppressed by the square of the resolved-sideband ratio $\Omega/\kappa_s$.
The corresponding occupancy is calculated by numerically solving the Born-Markov master equation (see Methods and Supplemental Material \cite{Supp}) and shown by the dashed blue line in Fig. \ref{figSOM} (a).

A photon that does enter the probe mode, corresponding to the state $\ket{ n_b  n_p  n_s}=\ket{010}$, can introduce noise by undergoing the non-resonant phonon-creating transition $\ket{010}\rightarrow\ket{101}$ (see Fig. \ref{Fig_Energy} (b)).
The resulting state can also resonantly transition to a two-phonon state, $\ket{101}\rightarrow\ket{210}$, as shown in Fig. \ref{Fig_Energy} (c).
Similarly to above, noise phonons from this process are suppressed by $\sim(\Omega/\kappa_p)^2$. 
Predicted phonon occupancies are shown in Fig. \ref{figSOM} (a) and (b).

\subsection*{Zero-point quadratic coupling rate required to fully test Bassi et al.'s lower bound}

The linearised quadratic part of the optomechanical interaction Hamiltonian is $H_{\rm int}^{(2)}=\bar n_{\rm cav}^{1/2}g_0^{(2)}(a^{\dagger}+a)(2b^{\dagger}b+b^{\dagger}b^{\dagger}+bb)$ \cite{Hauer_nondemolition_2018}. 
The term proportional to $b^{\dagger}b$ yields a per-phonon optical resonance frequency shift of $2\bar n_{\rm cav}^{1/2}g_0^{(2)}$, which is the signature of a collapse-induced phonon.
A random fluctuation $\delta \omega$ in the frequency of the probe can imitate a signal if it is larger or equal to this frequency shift, and sustained over a time comparable to the phonon lifetime $\Gamma^{-1}$. 
For shot noise limited probe, the probability of a fluctuation larger than $2\bar n_{\rm cav}^{1/2}g_0^{(2)}$ is given by an error function of a Gaussian distribution
\begin{equation}
p(\delta \omega)=\bigg(\frac{1}{\sqrt{2\pi \sigma^2}}\int^\infty_{\delta \omega}e^{-\omega^2/2 \sigma^2}d\omega\bigg),
\label{eq:Gauss}
\end{equation}
with standard deviation of
$\sigma\approx\kappa/\sqrt{N}$, where $N$ is
the number of photons interacting with a phonon within the mechanical lifetime $\Gamma^{-1}$, and is related to the average intracavity photon number via $N=\bar n_{\rm cav} \cdot \kappa/\Gamma$ for a continuous measurement. 
The rate of spurious signals due to such fluctuations is $R_{\delta\omega}=\Gamma p(\delta\omega)$.
To test a collapse-induced phonon flux of $\dot n_{\rm c}$, we require $\dot n_{\rm c}\geq R_{\rm \delta\omega}$.
From Eq. \ref{eq:Gauss} we find that, to fully exclude Bassi {\it et. al's} lower bound using the photonic-phononic crystal considered in the protocol above \cite{MacCabe_Ultralong_2019}, requires $\bar n_{\rm cav}^{1/2}g_0^{(2)}\gtrsim 3.5\sigma$.
Assuming an average intracavity photon number of $\bar n_{\rm cav}=10^2$, with $\kappa/2\pi=575$~MHz \cite{MacCabe_Ultralong_2019}, leads to the condition $g_0^{(2)}\gtrsim3.5\sqrt{\kappa\Gamma}\bar n_{\rm cav}^{-3/2}\gtrsim 2\pi\cdot 28$~Hz. 

The term proportional to $b^{\dagger}b^{\dagger}a$ converts a probe photon to two phonons, potentially imitating a collapse-signature. 
However, the shift induced by two phonons is $4\bar n_{\rm cav}^{1/2}g_0^{(2)}$ and can be clearly distinguished from the collapse-induced shift caused by one phonon.
Therefore, two-phonon creation can only imitate a collapse signal if it coincides with a frequency fluctuation of the probe mode $-\delta \omega\geq2 n_{\rm cav}^{1/2}g_0^{(2)}$, sustained at least over the two-phonon lifetime $(\sqrt2\Gamma)^{-1}$. The low probability of such a fluctuation, together with suppression on the order of $(2\Omega/\kappa)^2$ due to the non-resonant nature of the interaction, make this source of noise negligible.

\section*{Data availability}
The data that support the findings of this study are available within the paper and its Supplemental Material.
Codes for the numerical simulations are available on request from the corresponding author.

\section*{Author contributions}
WPB provided overall leadership for the project. SF and WPB conceptualized the idea. SF, SB, MZ and KK developed the theoretical model. SF and SB performed numerical simulations.  All co-authors contributed in the development of the manuscript which was drafted by SF and WPB.

\section*{Acknowledgements}
The authors thank Gerard Milburn, Nathan McMahon and James Bennett for helpful discussions, and Nicolas Mauranyapin for preparing Figure 1. We also acknowledge funding by Australian Research Council grants (EQUS, CE170100009, DE180101443) and the European Union Horizon 2020 research and innovation programme under the Marie Sklodowska-Curie grant agreement No 663830.

\section*{Additional information}

{\bf Competing interests:} The Authors declare no competing interests.

\newpage
\footnotesize \renewcommand{\refname}{\vspace*{-30pt}}  
\bibliographystyle{apsrev4-2} 
\bibliography{bib_v4}

\begin{thebibliography}{83}
\bibitem [{tbp()}]{tbp}%
  \BibitemOpen
  \href@noop {} {\bibinfo {title} {Calculations for other experiments to be
  published elsewhere}}\BibitemShut {NoStop}%
\end{thebibliography}


\end{document}


\title{Nanomechanical test of quantum linearity: Supplemental Material}

\author{Stefan Forstner$^1$, Magdalena Zych$^1$, Sahar Basiri-Esfahani$^2$, Kiran E. Khosla$^3$, and Warwick P. Bowen$^1$}
\address{$^1$Australian Centre for Engineered Quantum Systems, School of Mathematics and Physics, The University of Queensland, Australia}
\address{$^2$Department of Physics, Swansea University, Singleton Park, Swansea SA2 8PP, Wales, United Kingdom}
\address{$^3$QOLS, Blackett Laboratory, Imperial College London, United Kingdom}
\date{\today} 

\maketitle

 \section*{Supplementary Note 1. Signal rates from spontaneous collapse}
\label{sec:collapserates}
While there exists a plethora of collapse models \cite{Bassi_Review_2012,Bassi_Gravitational_2017}, they can all be formulated in terms of a stochastic nonlinear modification to Schr\"odinger's equation of the general form \cite{Ferialdi_Dissipative_2012,Adler_stochastic_2001}: 
\begin{equation}
\frac{d\Psi}{dt} = \bigg[-\frac{i}{\hbar}H + \sqrt{\lambda}A\xi(t)-\frac{\lambda}{2}(A^{\dagger}A+A^2) \bigg]\Psi,
\label{stochastic} 
\end{equation}
where the operator $H$ is related to the standard Hamiltonian of the system, $\xi(t)$ is defined in terms of an increment $dW(t)$ of a stochastic Wiener process $W(t)$ through
$\xi(t) dt = dW(t)$; 
$\lambda$ is a coupling that sets the strength of the collapse and $A$ is the reduction operator, which is specific to the particular realization of the collapse mechanism. The requirements of norm-preservation of the state evolution and {of} no-superluminal signalling make these collapse models the only mathematically consistent, phenomenological modifications against which quantum theory can be tested in this context \cite{Ferialdi_Dissipative_2012}.

In the following we give expressions for the  
decoherence rates due to two commonly studied collapse mechanisms. Collapse models 
differ in the properties and nature of the mechanisms purported to cause the collapse and can thus be classified according to the basis in which decoherence occurs, the mathematical properties of the stochastic mechanism and whether the mechanism has a quantum mechanical origin, or is due to a modification to Sch\"odinger's equation from a deeper-level theory. In the models discussed here decoherence acts in the position basis with Gaussian correlations in space, see \cite{Bassi_Review_2012} and references therein. 
A generic expression for the diffusion term $\sqrt{\lambda}A\xi(t)dt$ in these models is $\sqrt{\lambda}\int d\vec x \left(\rho(\vec x)-\langle{\rho(x)}\rangle\right) dW(\vec x,t)$ with  $\rho(\vec x)$ being the mass density, $W(\vec x,t)$ the ensemble of Wiener processes (for different points $\vec x$ in space) which can in general be correlated. The noise correlation length $r_c$ and the coupling rate $\lambda$ to the noise field  are the model parameters. The mass density typically takes the form $\rho(\vec x) = \sum_jm_j\int dy G(y-x)\psi_j^\dagger(y)\psi_j(y)$ with $G(y)$ a  ``smearing function'' and $\psi_j^\dagger(y), \psi_j(y)$ creation and annihilation operators of the particle with mass $m_j$. 
The stochastic Sch\"odinger equation Eq. \eqref{stochastic} and the above form of the diffusion term result in a master equation for the density operator $\hat\rho$ where the off-diagonal elements $\vec x', \vec x''$ evolve as $\frac{\partial}{\partial t}\bra{\vec x'} \hat\rho \ket{\vec x''} = -\tilde{D}( x',  x'')\bra{\vec x'} \hat\rho \ket{\vec x''}$ where 
 $\tilde{D}(\vec x', \vec x'') = \sum_{i,j}\frac{\lambda}{2}\left[G(\vec x_i' -\vec x_j')+G(\vec x_i'' -\vec x_j'')-2G(\vec x_i' - \vec x_j'')\right]$ 
and describes the rate at which the  spatial coherence is suppressed. Taking as an example the typically used Gaussian smearing function of the mass density $G(\vec x)=(4\pi r^2_c)^{-3/2}e^{-\vec x^2/4r_c^2}$, the decoherence rate reads $\tilde{D}(\vec x', \vec x'')=\lambda\cdot(4\pi r^2_c)^{-3/2}\cdot(1-e^{-|\vec x'-\vec x''|^2/4 r^2_c})$. We note that this is exactly the one-particle decoherence rate of the Ghirardi-Rimini-Weber (GRW) \cite{Ghirardi_Unified_1986} and in the Continuous Spontaneous Localization (CSL) models \cite{Pearle_Localization_1989,Ghirardi_Markov_1990}. Furthermore, for small superpositions sizes (as compared to the correlation length $r_c$) $|\vec x'-\vec x''|\ll r_c$ the decoherence rate becomes approximately $\tilde{D}(\vec x', \vec x'')\propto |\vec x'-\vec x''|^2/4 r^2_c$. In this limit one obtains as expected that the collapse rate increases with the size of superposition and the corresponding Lindblad term in the master equation takes a simple form $\propto - [\vec x, [\vec x, \hat\rho]]$. 

Position-basis decoherence yields localization of macroscopic superposition states. The same Lindblad term $-\tilde{D}[\vec x,[\vec x,  \hat\rho]]$ results in momentum diffusion of the system (this can be seen as a direct consequence of the Heisenberg Uncertainty 
Principle and the fact that the above decoherence process can equivalently be seen as a reduction of the position uncertainty of the system). In turn, the momentum diffusion can be interpreted as heating with the collapse-induced heating rate being directly proportional to the decoherence rate $\tilde{D}$. 
This allows testing the collapse models also by monitoring spontaneous heating of even isolated systems. 
For a quantum oscillator the heating leads to an increase of the phonon number expectation value. 
The equivalent Lindblad term reads $-\tilde{D}(b)[ b,[ b, \hat\rho]]$, with $\tilde{D}(b)=x_0^2\tilde{D}(\vec x', \vec x'')$, where $x_0$ is the zero-point motion (or equivalently, for spatial superpositions of massive particles, the superposition size $\Delta x$). 

The most studied collapse model is the 
CSL model \cite{Pearle_Localization_1989,Ghirardi_Markov_1990}. 
It considers second-quantized (albeit non-relativistic) indistinguishable particles where the collapse occurs in the particle number (Fock) basis. The key consequence of the indistinguishability of the particles is that for multi-particle systems the model predicts quadratic dependence of the decoherence rate on the number of particles that are within the cutoff distance $r_c$. For a comparison, the GRW model postulates discrete in time collapse events of the wave-function of individual (and also distinguishable) particles in the positions basis which yields linear dependence of the decoherence rate on the particle number \cite{Ghirardi_Unified_1986}.
The stochastic process in the CSL model is introduced in terms of a time-dependent Wiener noise at each point in space, coupling to mass-density smeared over some length scale $r_c$. As CSL is linear in the coupling rate $\lambda_c$ of matter to the collapse noise field, we define a dimensionless decoherence operator $D$ by $\tilde{D}_{\rm csl}(b)=\lambda_c D$.
%
For an oscillator with mass density $\rho(\vec x)$ and direction of motion along the $z$-axis, the decoherence operator in the CSL-model reads \cite{Nimmrichter_Optomechanical_2014,Vinante_Cantilever_2015}

\begin{equation}
D_{\rm csl}=\frac{(4\pi)^{3/2} r_c^3 x_0^2}{u^2}\int \frac{d^3\vec{k}}{(2\pi)^3}k^2_z e^{-k^2r_c^2}\vert\tilde{\rho}(\vec{k})\vert^2,
\end{equation}
where $\tilde{\rho}(\vec{k})=\int d^3r\rho(\vec{r})e^{-i\vec{k}\cdot\vec{x}}$ is the Fourier transform of the mass density and $u=1.66\cdot 10^{-27}$~kg is the atomic mass unit.

CSL in its original form predicts infinite energy increase as time goes to infinity \cite{Ghirardi_Markov_1990,Bassi_Dynamical_2003,Bassi_energy_2005}. This problem can be solved by postulating a finite temperature of the noise \cite{Smirne_Dissipative_2015}. Furthermore, the model assumes a white noise spectrum, which cannot be identified with any physical origin \cite{Ferialdi_Dissipative_2012}. In order to generalise the model to become compatible with relativity and with observations, the noise field should have a more general, i.e.~non-white spectrum. In such a case, however,  the model becomes non-Markovian~\cite{Adler_Nonwhite_2007, Adler_Nonwhite_2008}. While such models are difficult to study in full generality, it has been demonstrated~\cite{Adler_Nonwhite_2007, Adler_Nonwhite_2008} that to lowest order in $\lambda$ the qualitative features  of the model are the same as for the white-noise model.  A common assumption that helps to lift ambiguities in defining the model is that the  field underlying the collapse process has a cosmological origin. This allows one to introduce a high-frequency cutoff of the order of $\Omega_{\rm csl}/2\pi\approx 10^{10}-10^{11}$ Hz \cite{Bassi_Breaking_2010,Smirne_Dissipative_2015} which ensure the collapse rate is essentially as in a white-noise CSL model, but which changes the relaxation behaviour: in the coloured-noise model the system in the limit of long times thermalises to the temperature of the noise field, while in the white-noise model the system energy keeps growing with time \cite{Pearle_Collapse_1996}.

$D$ can be calculated analytically for simple geometries of composite test-systems \cite{Nimmrichter_Optomechanical_2014,Vinante_Cantilever_2015}.
For a sphere of radius $R$  the decoherence operator reads

\begin{equation}
D_{\rm sphere}=\frac{14\pi^2 R^2 r_c^2\rho^2 x_0^2}{3 u^2} \big(1-\frac{2r_c^2}{R^2}+e^{-R^2/r_c^2}\big(1+\frac{2r_c^2}{R^2}\big)\big),
\end{equation} 
and for the case of a cuboid with constant density $\rho$ and sidelengths $L_1$, $L_2$ and $L_3$, where $L_3$ is the direction of motion 
\begin{equation}
D_{\rm cuboid}= \frac{32 r_c^4\rho^2 x_0^2}{u^2}\big(1-e^{-\frac{L_3^2}{4r_c^2}}\big) \big(e^{-\frac{L_2^2}{4r_c^2}}-\frac{\sqrt{\pi}L_2}{2r_c}{\rm Erf}(\frac{L_2}{2r_c}) -1\big)  \big(e^{-\frac{L_1^2}{4r_c^2}}-\frac{\sqrt{\pi}L_1}{2r_c}{\rm Erf}(\frac{L_1}{2r_c}) -1\big).
\label{eq:dcuboid}
\end{equation}
For our proposed experiment, we estimate length, height and width of the photonic crystal beam to $L_1,L_2,L_3=1.21,0.22,0.22~\mu$m, respectively, reproducing the effective motional mass of the relevant mechanical mode of  136~fg \cite{MacCabe_Ultralong_2019}, where we used the density of silicon, $\rho=2.33\cdot10^{3}$~kg/m$^3$.  We choose the cuboid approximation for the modeshape following \cite{Vinante_Cantilever_2015}. From Eq. \eqref{eq:dcuboid} we find $D=5.1\cdot10^{5}$.

For the models where the collapse is assumed to have a gravitational origin, two main types of theories can be distinguished: where decoherence arises due to an intrinsic uncertainty in the local value of the gravitational field \cite{Diosi_Gravitation_1984,Diosi_Models_1989} or, equivalently, gravitational self-interaction \cite{Penrose_Reduction_1996}; and where decoherence is a consequence of the assumption that gravity is fundamentally a classical channel \cite{Kafri_Classical_2014,Khosla_Classical_2018,Altamirano_Pairwise_2018}. In both cases for small superposition size, the resulting effect has the same general form as the corresponding regime of the CSL and GRW models: For an oscillator it is proportional to the square of the zero point motion, for spatial superpositions of massive particles the effect is proportional the square of the superposition size.  Gravity-based decoherence has also been described within the framework of CSL in \cite{Ghirardi_gravity_1990}.

In the Di\'osi-Penrose (DP) model the decoherence rate is quantified by gravitational potential evaluated between superposed amplitudes of the system: 
$\frac{G}{2\hbar}[U(XX)+U(YY)-2U(XY)]$, where $U(XY)=-G\int d^3r\int d^3r'\frac{\rho_X(\vec x)\rho_Y(\vec x')}{|\vec x-\vec x'|}$ is the gravitational interaction between mass-densities $\rho_X, \rho_Y$ associated with the superposed configurations $X, Y$ \cite{Diosi_bulk_2014}, with $G=6.67\cdot10^{-11}$~m$^3$kg$^{-1}$s$^{-2}$ being Newton's gravitational constant. For point particles the above expression gives divergent decoherence rate and thus a short-distance cutoff $r_{\rm DP}$ is needed. The decoherence operator reads \cite{Nimmrichter_Optomechanical_2014}

\begin{equation}
\tilde{D}_{\rm DP}=\frac{x_0^2 G}{6\sqrt{\pi}\hbar}\big(\frac{a}{r_{\rm DP}}\big)m\rho(\vec x),
\label{eq:DPheating}
\end{equation}
where $a$ is the lattice constant of the composite object.
Comparing the heating rate expected from Eq. \eqref{eq:DPheating} to measurement-induced spurious phonons, which constitute the strongest noise source in our proposed experiment (see main text), we find that short-distance cutoffs up to $r_{\rm DP}\approx 3.9$~fm can be excluded. For a discussion of experimental tests of classical channel gravity \cite{Kafri_Classical_2014,Khosla_Classical_2018,Altamirano_Pairwise_2018}, see \cite{tbp}.



\section*{Supplementary Note 2. Calculation details: efficiency and noise levels of the optomechanical system}
\label{sec:optomechanics}
The optomechanical system is probed by a strongly attenuated coherent source, such as described in \cite{Kessler_Laser_2011}, which can be stabilized to a sub-Hz linewidth $\kappa_L$. Because $\kappa_L$ is much smaller than $\kappa_s$, $\kappa_p$, and $\kappa_f$, which are the linewidths of the {\it probe mode}, the {\it signal mode}, and the filter cavity, respectively, the field $a_{\rm in}$ of the incoming probe laser is well approximated with a $\delta$ - function: $a_{\rm in}(\omega)=a_{\rm in}\delta(\omega_L)$, where $\omega_{\rm L}$ is the laser frequency. We set $\omega_L=\omega_p$ to maximize coupling into the probe mode $a_p$. 
The mechanical decay rate $\Gamma$, as well as the collapse rate are slow compared to timescales of the optomechanical interaction:  $D,\Gamma \ll g_0,\kappa_{p/s}$, where $g_0$ is the single photon optomechanical coupling rate \cite{Aspelmeyer_Review_2014}. 
In this limit, the conversion dynamics can be modelled by the reduced master equation:
\begin{equation}
\frac{d\hat \rho}{dt}=-\frac{i}{\hbar}\big[H_{\rm int},\hat\rho\big] +\kappa_p \mathcal{D}[a_p]\hat\rho + \kappa_s \mathcal{D}[a_s]\hat\rho,
\label{eq:reducedME}
\end{equation}
where $\hat \rho$ is the density matrix, the operators $a_p$ and $a_s$ correspond to annihilation operators for the optical probe- and signal-modes, respectively, and $H_{\rm int}$ is the interaction Hamiltonian given in the main text.

\begin{figure}[h!]

  \includegraphics[width=\textwidth]{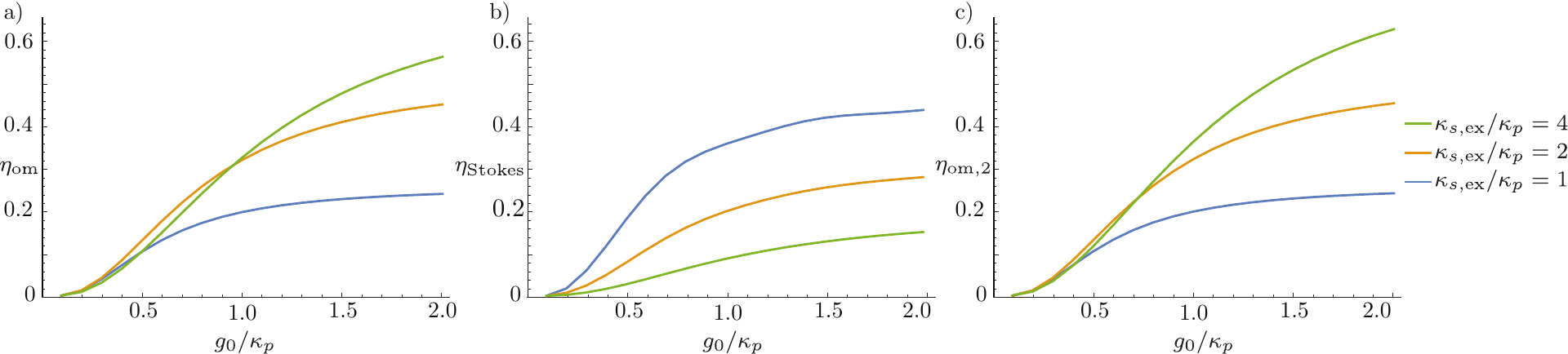}
\caption{Efficiencies of optomechanical transitions.
a) Cumulative probability for the output of a signal photon due to anti-Stokes scattering from the initial state $\ket{ n_bn_p n_s}=\ket{110}$ as a function of $g_0/\kappa_p$, for $\kappa_{s,\rm ext}/\kappa_p=1,2,$ and $4$.
b) Efficiency of spurious Stokes scattering, corresponding to the phonon number occupancy expectation value $\eta_{\rm Stokes}=\langle b^{\dagger}(t)b(t)\rangle$ after an incident photon in the signal mode, for times $\Gamma^{-1}\gg t \gg \kappa_p^{-1},\kappa_s^{-1}$.
c) Cumulative probability for the output of a signal photon due to anti-Stokes scattering from the two-phonon state $\ket{n_b n_p n_s}=\ket{210}$.
}

\label{fig_hOM}
\end{figure}

{\it Resonant anti-Stokes scattering.} 
If a phonon is introduced into the ground-state cooled oscillator, and a probe pulse is incident within the lifetime of the mechanical excitation, the system is in the initial state $\ket{ n_b n_p  n_s}=\ket{110}$. 
The optomechanical conversion efficiency $\eta_{\rm om}$ is the probability of one photon in the probe mode $a_p$ scattering with one phonon in the mechanical resonator, creating a photon in the signal mode $a_s$ ($\ket{110}\rightarrow\ket{001}$) via a anti-Stokes Raman process, and this photon being outcoupled to create a signal photon at frequency $\omega_s$. 
$\eta_{\rm om}$ is obtained by numerically solving Eq. (\ref{eq:reducedME}) and time-integrating over the emission from the signal mode, $\eta_{\rm om}=\kappa_{s,\rm ex}\int_0^{\infty}\langle a_s^{\dagger}(t)a_s(t)\rangle dt$, where $\kappa_{s,\rm ex}=\kappa_s-\kappa_{s,0}$ is the external decay rate of the signal mode due to coupling, with $\kappa_{s,0}$ and $\kappa_s$ its intrinsic and loaded decay rates, respectively.
Supplementary Figure \ref{fig_hOM} (a) shows $\eta_{\rm om}$ as a function of the effective coupling strength $g_0/\kappa_p$ for critical coupling of the probe mode $\kappa_{p,0}=\kappa_{p,\rm ex}$ and equal intrinsic couplings $\kappa_{s,0}=\kappa_{p,0}$, for critically ($\kappa_{s,0}=\kappa_{s,\rm ex}$) and overcoupled ($\kappa_{s,\rm ex}/\kappa_p=2$ and $4$) signal mode. 
The efficiency $\eta_{\rm om}$ is sensitive to the ratio $g_0/\kappa_p$, as $g_0$ sets the rate for anti-Stokes scattering, and $\kappa_p$ sets the rate for the competing optical decay.
Furthermore, it depends on the ratio $\kappa_{s,\rm ex}/\kappa_p$ as higher values favour outcoupling of the signal photon. 
For our proposed experiment, $g_0=\kappa_p$ and $\kappa_{s,\rm ex}/\kappa_p=1.8$.

The probability of a phonon in the mechanical resonator translating to a coincidence count, imitating a signal, is given by $\eta=\eta_p\eta_{\rm om}\eta_{\rm f}\eta_{\chi}\eta_d=1.1\cdot10^{-3}$, where $\eta_p$ is the probability of a photon entering the probe mode during the mechanical excitation lifetime, $\eta_f=0.56$ is the transduction efficiency through the filter for a signal photon at frequency $\omega_s$, $\eta_{\chi}=0.95$ and $\eta_d=0.64$ are downconversion and coincidence detection efficiencies, respectively. These efficiencies are analysed in more detail in the following paragraphs. As the rate of phonons created in the resonator due to spontaneous collapse is given by the collapse rate $\lambda_cD$, the rate of registered collapse signatures is $R_c=\lambda_cD\eta=5.5\cdot 10^{2}\lambda_c$.

{\it Probe field occupancy.}
The average number of photons encountered by one phonon is given by
\begin{equation}
\bar n_{\rm ph}=\int_0^\infty e^{-\Gamma t}\kappa_p\bar n_p dt,
\label{eq:etap}
\end{equation}
where $\bar n_p$ is the average photon occupancy of the probe mode and $e^{-\Gamma t}$ is the phonon occupancy at time $t$ after one phonon is created in the mechanical resonator at time $t=0$. 
In the limit $\bar n_p \ll \Gamma/\kappa_p$, Eq. (\ref{eq:etap}) also quantifies the probability $\eta_p$ of a phonon in the mechanical resonator encountering a probe photon, $\eta_p=\bar n_{\rm ph}$. 
Furthermore, in this limit, $\eta_p$ asymptotes to $\eta_p=\bar n_p\kappa_p/\Gamma$. 
In our protocol, the probe laser power is adjusted so that $\eta_p=0.01$, corresponding to an average of 0.01 photons per mechanical oscillator lifetime.
In the steady state, the average intracavity photon number is given by 
$\bar n_p=4\kappa_{p,\rm ex}\bar n_{\rm in}/\kappa_p^2$ \cite{Aspelmeyer_Review_2014},
and hence, to achieve a given $\eta_p$ the input field occupancy is adjusted to $\bar n_{\rm in}(\eta_p)=(\eta_p\Gamma\kappa_p)/(4\kappa_{p,\rm ex})$. 
Because the signal is proportional to $\eta_p$, and the noise background from measurement-induced noise phonons is proportional to $\eta_p^2$, the input field occupancy gives a handle to lower the noise at the cost of longer measurement time, or vice versa.

{\it Transmission through the filter cavity.} 
The efficiency  $\eta_f$ of the signal passing through the filter, assuming equal input- and output coupling strengths $\kappa_{f,\rm ex}$, is given by \cite{Hecht_Optics_2002}
\begin{equation}
\eta_f=\big(1-\frac{\kappa_{f,0}}{\kappa_f}\big)^2,
\end{equation}
where $\kappa_{f,0}$ is the intrinsic filter linewidth and $\kappa_f=\kappa_{f,0}+\kappa_{f,\rm in}+\kappa_{f,\rm out}$ is the loaded filter linewidth, with $\kappa_{f,\rm in}$ and $\kappa_{f,\rm out}$ the in- and output coupling, respectively.
Using a typical a laser stabilisation filter cavity \cite{Kessler_Laser_2011},  $\kappa_{f,0}=30$~kHz, and overcoupling both at the input and output $\kappa_{f,\rm in}=\kappa_{f,\rm out}=1.5\cdot\kappa_{f,0}$, a transmission efficiency of $\eta_f=0.56$ is achieved.

{\it Nonlinear downconversion.} 
After separation from probe light at frequency $\omega_p$, signal photons at frequency $\omega_s$ are downconverted to photon pairs in a nonlinear medium.
In order to minimize detector dark counts, we propose a nonlinear conversion process to convert signal photons to pairs. A bright classical pump beam with electric field amplitude $E$ is coupled into a medium exhibiting a third order $\chi^{(3)}$ optical nonlinearity. This yields an effective second order interaction 
$H_{\chi,\rm eff}=\gamma E a_{f,\rm out}d_1^{\dagger}d_2^{\dagger}+\gamma^* E a_{f,\rm out}^{\dagger}d_1d_2$, where $\gamma$ is the nonlinear coupling strength \cite{Langford_Conversion_2011}, $a_{f,\rm out}$ is the mode of the signal transmitted through cavity and filter, and $d_{1/2}$ are the modes coupled to the detectors. 
Given the input state $\ket{n_{f,\rm out}n_{d1}n_{d2}}=\ket{100}$, the time evolution in the nonlinear medium is \cite{Langford_Conversion_2011}
\begin{equation}
\ket{\Psi(t)}=\cos{(\abs{\gamma}t/\hbar)}\ket{100}+i \sin{(\abs{\gamma}t/\hbar)}\ket{011}.
\end{equation}
 
By setting the length of the nonlinear medium to $L=\frac{1}{2}\pi\hbar c_n\abs{\gamma}^{-1}$, where $c_n$ is the speed of sound in the medium, the output state is $\ket{\Psi(t_{\rm final})}=\ket{011}$, corresponding to a photon in each detector mode $d_{1/2}$.
It has been shown that this conversion can be performed with near-unit efficiency \cite{Langford_Conversion_2011}, hence we assume $\eta_\chi=0.95$.

{\it Coincidence detection.}  
Assuming a detection efficiency of $80\%$ for a single detector \cite{Photonspot_private}, the efficiency for coincidence detection is $\eta_d=(0.80)^2=0.64$.  
The coincidence dark count rate is $R_{\rm coincidence}=R_{\rm d}^2 \cdot \tau_c$, where $R_{\rm d}$ is the dark count rate of a single detector and $\tau_c$ is the coincidence timing resolution (time jitter). 
This allows a suppression of dark counts with the square of the single-detector dark count rate. For $R_{\rm d}=3.5$~Hz and $\tau_c=30$~ps \cite{Photonspot_private}, the predicted coincidence dark count rate is $R_{\rm coincidence}=3.7\cdot10^{-10}$~s$^{-1}$. 
 
{\it Probe photons leaking through the system.}
There are two ways in which probe photons can leak through the system and potentially imitate a collapse signature.
Firstly, optomechanical conversion processes can create photons at frequency $\omega_p$, or at a frequency reduced by multiple integers $n$ of the mechanical resonance frequency, $\omega_p-n\Omega$.
Discussions of these processes are included in the following paragraphs on noise phonons.
The amplitudes of these processes are negligible due to strong suppression (see main text).
Secondly, measurement noise is introduced by probe photons transmitted through the filter cavity and downconverted to a pair of photons in the nonlinear medium, imitating a signal. 
The probability of a probe photon with detuning $\Delta=\omega_s-\omega_p=\Omega$ transmitting through both the near-critically coupled optomechanical system and a subsequent filter cavity of linewidth $\kappa_f$ and free spectral range $\omega_{\rm fsr}$ is given by \cite{Hecht_Optics_2002}

\begin{equation}
p_f(\Omega)=\eta_f\big[ 1 + (\frac{4}{\kappa_f})^2 \cdot {\rm sin}^2{(\frac{\pi\Delta}{\omega_{\rm fsr}})}\big]^{-1}.
\label{eqFilter}
\end{equation}
For the frequency of the proposed mechanical resonator of $\Omega/2\pi=5.3$~GHz \cite{Chan_Nanomechanical_2011}, a finesse of $3.16\cdot10^{5}$ as in a standard laser stabilisation reference cavity \cite{Kessler_Laser_2011} and a cavity length $L$ of one centimeter, we find $\omega_{\rm fsr}/2\pi=c/2L=15$~GHz and $p_f=3.5\cdot 10^{-10}$.
A noise photon is then downconverted to a photon pair and registered as a coincidence count with efficiency $\eta_\chi\eta_d$. 
As the rate of incoming probe photons coupled into the probe mode in the steady state is $4\kappa_{p,\rm ex}\bar n_{\rm in}/\kappa_{p}=\bar n_p\kappa_p=\eta_p\Gamma$, the rate of coincidence counts due to noise photons, imitating a signal, is 
\begin{equation}
R_{\rm phot}=\eta_p\Gamma p_f \eta_\chi\eta_d=1.4\cdot 10^{-12}\,{\rm s}^{-1}.
\label{eqRf}
\end{equation}


{\it Noise phonons from direct occupation of the signal mode.}
A photon from the coherent probe laser can create a phonon by coupling into the signal mode $a_s$ instead of the probe mode $a_p$, which results in a direct occupation of the signal mode $\ket{ n_b  n_p  n_s}=\ket{001}$, as shown in Fig. 3 (a) in the main text. 
This process is suppressed due to the small spatiotemporal overlap $\Theta$ of the signal mode with the coherent laser beam, $\Theta=\kappa_s^2/(\kappa_s^2+\Omega^2)\approx(\kappa_s/\Omega)^2=3.2\cdot 10^{-5}$, where $\kappa_s/2\pi=30$~MHz is the loaded decay rate of the signal mode.
If the photon is outcoupled, it has a frequency of $\omega_p$ due to energy conservation and can be efficiently filtered from a signal at frequency $\omega_s$.
However, a scattering process to the probe mode can create a noise phonon in mode $b$, imitating a decoherence-signature.
The efficiency of this Stokes scattering process is given by the probability of the initial state $\ket{001}$ causing the mechanical oscillator to be in the excited state, which equals the phonon occupancy after optical decay, $\eta_{\rm Stokes}=\langle b^{\dagger}(t)b(t)\rangle$, for times $t\gg\kappa_p^{-1},\kappa_{s}^{-1}$.
$\eta_{\rm Stokes}$ is shown as a function of $g_0/\kappa_p$, for $\kappa_{s,\rm ex}/\kappa_p =1,2,$ and $4$ in Supplementary Figure \ref{fig_hOM} (b).
Higher values of $\kappa_{s,\rm ex}$ correspond to lower probabilities for this type of noise, as the decay process of rate $\kappa_{s,\rm ex}$ competes with the Stokes scattering of rate $g_0$.
For the proposed experiment with $\kappa_{s,\rm ex}/\kappa_p=1.8$ we find $\eta_{\rm Stokes}=0.17$.
The probability of phonon creation, due to this process, 
at time $t_0$ after incidence of the probe photon is $p_{\rm direct}=(\kappa_p/\kappa_{p,\rm ex})\cdot\Theta\eta_{\rm Stokes}=2.7\cdot10^{-5}$.

{\it Noise phonons from counterrotating optomechanical processes.}
Another mechanism for noise phonon creation is given by probe photons anti-Stokes scattering into the signal mode, corresponding to the resonantly suppressed (counterrotating) transition   $b^{\dagger}a_p a_s^{\dagger}e^{-2i\Omega t}\ket{010}\rightarrow\ket{101}$. 
The resulting state contains a photon in the signal mode as well as a phonon in the mechanical resonator, which could in principle both imitate a signal. 
However, the outcoupled photon has a frequency of $\omega_p-\Omega$ and can, therefore, be filtered from the signal at frequency $\omega_s$.
The probability of phonon creation due to this process, caused by a single incident probe photon, is obtained by numerically solving Eq. (\ref{eq:reducedME}), is $p_{\rm counterrot,1}=1.7\cdot 10^{-5}$, where $p_{\rm counterrot,1}=\bra{001}\hat \rho(t\gg\kappa_p^{-1},\kappa_s^{-1}) \ket{100}+\bra{101}\hat \rho(t\gg\kappa_p^{-1},\kappa_s^{-1}) \ket{101}$, with $\hat \rho(t)$ the density matrix initialized in the state $\hat\rho(t=0)=\ket{010}\bra{010}$  (see Fig. 4 (a) and (b) in the main text).
Further, the state $\ket{101}$ can resonantly transition to a state with two phonons in the mechanical resonator and one photon in the probe mode: $b^{\dagger}a_p^{\dagger}a_s\ket{101}\rightarrow\ket{210}$.
The probability of one incident probe photon preparing the system in this two-phonon state $p_{\rm counterrot,2}=$\mbox{$\bra{002}\hat \rho(t\gg\kappa_{p}^{-1},\kappa_s^{-1}) \ket{200}$}$+\linebreak$\mbox{$\bra{012}\hat \rho(t\gg\kappa_p^{-1},\kappa_s^{-1}) \ket{210}$} (with $\hat\rho(t=0)=\ket{010}\bra{010}$) numerically obtained by solving Eq. (\ref{eq:reducedME}) (see Fig. 4 (b) in the main text), finding $p_{\rm counterrot,2}=1.6\cdot 10^{-6}$.
We find that the occupancies $p_{\rm counterrot,1}$ and $p_{\rm counterrot,2}$ scale with $(\kappa_p/\Omega)^2$ in the limit $g_0=\kappa_p$, with higher order transitions suppressed by $(\kappa_p/\Omega)^4$ and therefore negligible.

{\it Scattering of noise phonons to signal photons.}
A spurious signal photon at frequency $\omega_s$ is created if a noise phonon scatters with a second photon entering the probe mode within the lifetime of the mechanical excitation. 
For timescales of the mechanical excitation lifetime,  $t \sim \Gamma^{-1}\gg \kappa_{p}^{-1},\kappa_{s}^{-1}$, after incidence of a probe photon, it is convenient to define the `occupancy probabilities' of the one- and two-phonon states from the optomechanical processes described above: $p_{n_b=1}(t_0)=p_{\rm direct}+p_{\rm counterrot,1}$ and $p_{n_b=2}(t_0)=p_{\rm counterrot,2}$, where $t_0$ is long compared to timescales of the optical decay, so that all optomechanical conversions have concluded, but short compared to the mechanical decay ($\Gamma^{-1}\gg t_0\gg\kappa_{p}^{-1},\kappa_{s}^{-1}$). The time dynamics of these one- and two phonon occupancy probabilities are described by
\begin{equation}
p_{n_b=1}(t)=n_{n_b=1}(t_0) e^{-\Gamma t}+2p_{n_b=2}(t_0)\big( e^{-\Gamma t}-e^{-2\Gamma t} \big)
\end{equation} 
and 
\begin{equation}
p_{n_b=2}(t) =p_{n_b=2}(t_0)  e^{-2\Gamma t},
\end{equation}
respectively.
The probability of these measurement-induced phononic states encountering a second probe photon is given by $\eta_p\Gamma\int_0^\infty p_{n_b=1}(t) dt$ and $\eta_p\Gamma\int_0^\infty p_{n_b=2}(t) dt$, respectively. The conversion efficiency of the one- and two-phonon state to a spurious photon outcoupled from the cavity is then given by $\eta_{\rm om}$ and $ \eta_{\rm om,2}$, respectively, where latter is numerically calculated, and plotted as a function of $g_0/\kappa_p$ for different values of $\kappa_{s,\rm ex}/\kappa_p =1,2,$ and $4$, as shown in Supplementary Figure \ref{fig_hOM} (c).
The resulting probability of a noise phonon of frequency $\omega_s$ coupled out of the cavity, due to optomechanical processes, potentially imitating a signal, is
\begin{equation}
p_{\rm om}(t)= \eta_p\Gamma\int_0^t p_{n_b=1}(t)\eta_{\rm om}+ p_{n_b=2}(t)\eta_{\rm om,2} dt,
\label{eq:pom} 
\end{equation}
as shown in the main text in Fig. 4 (c). 
We find the asymptotic value $p_{\rm om}(t\rightarrow\infty)=8.4\cdot 10^{-8}$.
In analogy to Eq. (\ref{eqRf}), the rate of coincidence counts due to noise is given by $R_{\rm om}=\eta_p\Gamma p_{\rm om}(t\rightarrow\infty)\eta_f \eta_\chi\eta_d=1.9\cdot 10^{-10}$~s$^{-1}$.

{\it Noise phonons due to photoabsorptive heating.}
The number of noise phonons originating from absorption heating is estimated following \cite{Meenehan_millikelvin_2014,Ren_Crystal_2019}.
We compute the average phonon number excited by one probe photon in the mechanical resonator, $\bar n_{\rm abs,1}$, due to photoabsorption for time $t_{\rm abs}$, at which the oscillator is in thermal equilibrium with the material, but not yet with the environment  ( $\Gamma^{-1}\gg t_{\rm abs} \gg \gamma_{\rm THz}^{-1}$, with $\gamma_{\rm THz}$ the rate at which THz-frequency phonons radiate to the environment, see also Methods).
We approximate $\bar{n}_{\rm abs,1}$ by extrapolating from \cite{MacCabe_Ultralong_2019,Ren_Crystal_2019}.
In this case, an average intracavity photon number $ \bar n_{\rm cav}=1$ yields an average phonon number of $\bar n_{\rm abs}=10$ in the mechanical resonator.
$\bar n_{\rm cav}=1$ is equivalent to $\kappa/\Gamma$ photons passing through the cavity within the lifetime $\Gamma^{-1}$ of the mechanical resonator, creating in total 10 phonons.
The intrinsic optical decay is limited by photon absorption, $\kappa_0\approx\kappa_{\rm abs}$ and  $\gamma_{\rm THz} \approx \kappa_0/2\pi$, within the errors given. 
Therefore, from one photon we expect the induced occupancy $\bar{n}_{\rm abs,1}=10\frac{\Gamma}{\kappa}=10\cdot\frac{108\  \rm mHz/2\pi}{575\ \rm MHz/2\pi}\approx 1.9\cdot 10^{-9}$.


As with the optomechanical heating rates, a spurious signal will only occur if the phonons resulting from absorption heating interact with another probe photon. In analogy to Eq. (\ref{eq:pom}), the probability of a probe photon creating a spurious signal photon at frequency $\omega_s$ due to absorption heating is 
\begin{equation}
p_{\rm abs}(t)=\eta_p\Gamma \int_0^t\bar n_{\rm abs,1} e^{-\Gamma t'}\eta_{\rm om} dt'.
\label{eq:pabs}
\end{equation}
For the proposed setup we find $p_{\rm abs}(t\rightarrow\infty)=6.1\cdot 10^{-12}$ and  $R_{\rm abs}=1.4\cdot10^{-14}$~s$^{-1}$.

For the quadratic coupling approach, $\bar n_{\rm cav}=10^2$ (see main text), and the relation $\bar n_{\rm abs} \propto \bar n_{\rm cav}^{1/3}$ holds \cite{Meenehan_millikelvin_2014,Ren_Crystal_2019}, and we find an average intracavity phonon number of $\bar n_{\rm abs} \approx 5$. This corresponds to a noise phonon rate of $\dot n=\Gamma\bar n_{\rm abs} \approx 3$~s$^{-1}$, close to seven orders of magnitude higher than the lowest possible phonon flux expected from Bassi {\it et al.}'s proposal of $\dot n_c=5.1\cdot10^{-7}$~s$^{-1}$.

{\it Multiplexing.}
A photonic-phononic crystal, including suspension and phononic shield, requires an area of about 1000 $\mu$m$^2$ \cite{MacCabe_Ultralong_2019}. Thus it would be conceivable to fabricate a high number of them on a 4-inch-wafer with an area of $\sim 2\cdot 10^{9}$ $\mu$m$^2$. We assume $N\sim10^{4}$, allowing more than 99$\%$ of the wafer to be reserved for waveguide coupling, fabrication tolerances, etc. In principle, all these devices to may be connected to the same filter cavity, nonlinear medium, and detector pair, or to a small number of such elements.  

\section*{References}

\footnotesize \renewcommand{\refname}{\vspace*{-30pt}}  
\bibliographystyle{apsrev4-2} 

\bibliography{bib_Supp_v2}
